\documentclass[conference]{IEEEtran}
\usepackage{graphicx}
\usepackage[center]{caption}
\usepackage{subcaption}
\usepackage{cite}
\usepackage{dsfont}
\usepackage{color}
\usepackage{amsmath,amssymb,amsthm}
\usepackage{lettrine}
\newcommand\norm[1]{\left\lVert#1\right\rVert}
\usepackage{algorithm,algorithmic,multicol}

\newtheoremstyle{mystyle}
{}
{}
{\itshape}
{}
{\bfseries}
{.}
{ }
{}
\theoremstyle{mystyle}

\begin{document}
\title{Hardware Efficient Joint Radar-Communications with Hybrid Precoding and RF Chain Optimization}

\author{\IEEEauthorblockN{Aryan Kaushik, Christos Masouros, Fan Liu}
\IEEEauthorblockA{Department of Electronic and Electrical Engineering, University College London, United Kingdom. \\
Emails: \{a.kaushik, c.masouros, fan.liu\}@ucl.ac.uk}}

\maketitle

\begin{abstract}
In this paper, we aim to achieve energy efficient design with minimum hardware requirement for hybrid precoding, which enables a large number of antennas with minimal number of RF chains, and sub-arrayed multiple-input multiple-output (MIMO) radar based joint radar-communication (JRC) systems. A dynamic active RF chain selection mechanism is implemented in the baseband processing and the energy efficiency (EE) maximization problem is solved using fractional programming to obtain the optimal number of RF chains at the current channel state. Subsequently hybrid precoders are computed employing a sub-arrayed MIMO structure for EE maximization with weighted formulation of the communication and radar metrics, and the solution is based on alternating minimization. The simulation results show that the proposed method with minimum hardware achieves the best EE while maintaining the rate performance, and an efficient trade-off between sensing and communication.
\end{abstract}

\begin{IEEEkeywords}
EE maximization, joint radar-communications, hybrid precoding, RF chain optimization. 
\end{IEEEkeywords}

%
\IEEEpeerreviewmaketitle

\section{Introduction}
The sharing of spectral and hardware resources to deploy joint radar-communication (JRC) systems \cite{blissACCESS2017} is a promising solution in decongesting the limited radio frequency (RF) spectrum efficiently and achieve better functionality than existing individual sensing or wireless communication systems \cite{fanTCOM2020, hassanienSP2019, maSP2020}. 
A number of radar and communication coexistence scenarios involve millimeter wave (mmWave) frequencies above 30GHz 
\cite{zhangTVT2019, dokhanchiTAES2019}. Such bands are typically considered for short ranges because of high attenuation from physical barriers, however compact antenna arrays with high beamforming gains would also extend the coverage range\cite{mishraSP2019}. This paper makes use of mmWave channel, however the proposed method is parametric and independent of channel modeling. 

The wireless local area network (WLAN) standard, e.g., 802.11ad WLAN protocol addressed in \cite{kumariTVT2018, grossiTSP2018} operating at 60 GHz band, typically employs small number of antennas which can support short-range sensing only. The use of large scale multiple-input multiple-output (MIMO) antenna setup can overcome such issue in addition to compensating the high path loss associated with mmWave. It also provides more degrees of freedom to enable joint sensing and communication, e.g., \cite{fanWCL2017, cuiSPAWC2018, fanTSP2018} use the MIMO based JRC system where there are multiple radar users with multiple communication users. Most of existing literature discussed above uses fully digital precoding based MIMO JRC systems where the number of RF chains is same as the number of antennas, which leads to high hardware complexity and large power consumption. 

In order to reduce hardware complexity and power consumption with spatial multiplexing functionality, hybrid precoding can be implemented \cite{aryanIET2016, ayachTWC2014, aryanICC2019}. For the MIMO JRC system, we can consider 
the sub-arrayed MIMO radar \cite{sellathuraiTSP2012} to further save power consumption. As per \cite{sellathuraiTSP2012, hassanienTSP2010}, the sub-arrayed MIMO radar achieves a performance trade-off between phased-array radar with single RF chain and MIMO radar with same number of RF chains as the antennas. 
References \cite{fanTCOM2020, fanICASSP2019} consider hybrid precoding but with fixed number of RF chains, which are complex and power hungry components, and their aim is only achieving near optimal rate. Notably, for these JRC systems, hardware efficient solutions with low power consumption are not widely studied. Selecting only the required number of RF chains dynamically for current channel state can significantly reduce the power consumption and hardware complexity. Reference \cite{aryanTGCN2019} designs such framework for mmWave MIMO communication only systems.

In this paper, we implement a dynamic RF chain selection mechanism for hybrid precoding and sub-arrayed MIMO radar based JRC systems. We consider an energy efficiency (EE) maximization problem 
under radar and communication constraints. The problem becomes difficult due to non convexity of the cost function, which is 
solved using fractional programming to obtain the optimal number of RF chains. Subsequently we model the hybrid precoder design as weighted minimization problem where the optimal radar precoder and optimal fully digital precoder matrices are approached depending upon the weights on radar and communication metrics. This problem is solved using alternating minimization. The obtained optimal number of RF chains and hybrid precoder matrices are used to compute maximized EE, and the performance of the proposed method is evaluated by the simulation results. 

\emph{Notation:} 
$(.)^{T}$ stands for transpose, $(.)^{H}$ is complex conjugate transpose, $\norm{(.)}_F$ is Frobenius norm, tr(.) is trace, $|.|$ is determinant, $[.]_{kl}$ is the matrix entry at the $k$-th row and $l$-th column,
$\textbf{I}_{N}$ is $N$-size identity matrix, $\mathbb{C}$, $\mathbb{R}$ and $\mathbb{R}^+$ denote the sets of complex numbers, real numbers and positive real numbers, respectively, 
and $\mathbb{E}\{\cdot\}$ is the expectation operator. 

\section{JRC System with Hybrid Precoding}
\subsection{Channel Model}
We consider a mmWave downlink channel with $N_{\textrm{T}}$ antennas at the base station (BS), that transmits $N_\textrm{s}$ data streams towards $N_\textrm{R}$ users (UEs),
and multiple point-like targets towards which the BS steers beams in the sensing environment. 
The flat fading narrowband channel model with $N_\textrm{m}$ scattering multipaths is expressed as follows \cite{sohrabiJSTSP2016}:
\begin{equation}\label{eq:channel_model}
\mathbf{H} = \sqrt{\frac{N_\textrm{T}N_\textrm{R}}{N_\textrm{m}}} \sum_{l=1}^{N_{\textrm{m}}}  \alpha_{l} \mathbf{a}_{\textrm{R}}(\phi_{l}^{r}) \mathbf{a}_{\textrm{T}}(\phi_{l}^{t})^H,
\end{equation}
where $\alpha_{l}$ is the gain term, $\textbf{a}_{\textrm{T}}(\phi_{l}^{t})$ and $\textbf{a}_{\textrm{R}}(\phi_{l}^{r})$ denote the normalized transmit and receive array response vectors, respectively, with $\phi_{l}^{t}$ being the angles of departure and $\phi_{l}^{r}$ being the angles of arrival. We consider uniform linear array (ULA) setup, 
e.g., for a $N_{z}$-element ULA on $z$-axis, 
$\textbf{a}_{z}(\phi) = \frac{1}{\sqrt{N_{z}}}{[1, e^{j \frac{2 \pi}{\lambda}d\sin(\phi)}, ..., e^{j (N_{z}-1)\frac{2 \pi}{\lambda}d\sin(\phi)}]}^{T}$ \cite{balanis1997},
where $\lambda$ is the signal wavelength and $d\!=\!\!\lambda/2$ is the inter-element antenna spacing. 
We assume that channel is known to the BS.

\subsection{System Model} Fig. 1 shows block diagram of a MIMO JRC system with hybrid precoding. At the BS, the digital precoder unit is followed by $L_\textrm{T}$ RF chains and associated digital-to-analog converter (DAC) units where we assume full-bit resolution. Before transmission, the signal is processed by a network of phase shifters, i.e., analog precoder unit. 
The transmit symbol vector $\mathbf{s} \in \mathbb{C}^{N_\textrm{s} \times 1}$ at the BS is such that $\mathbb{E}\{\mathbf{s} \mathbf{s}^H\} = \mathbf{I}_{N_\textrm{s}}$. We consider the sub-arrayed MIMO radar \cite{sellathuraiTSP2012} connecting each RF chain to only a subset of the antennas, i.e., $N_\textrm{T}/L_\textrm{T}$, requiring less number of phase shifters. The analog preocder $\mathbf{F}_{\textrm{RF}}$ has diagonal entries $\textbf{f}_i \in \mathbb{C}^{N_\textrm{T}/L_\textrm{T} \times 1}\, \forall i = 1,..,L_\textrm{T}$, where $\textbf{f}_i$ are the values of the phase shifters at the $i$-th sub array containing constant-modulus entries, i.e., the elements have unit modulus and continuous phase. 
The matrix $\mathbf{F}_{\textrm{BB}}$ 
represents the baseband precoder. The power constraint for downlink communication is satisfied by $\lVert\mathbf{F}_\textrm{RF}\mathbf{F}_\textrm{BB}\rVert_F^2$ = $P_{\textrm{max}}$, where $P_{\textrm{max}}$ is the maximum allocated power. The receiver output signal after combiner processing is expressed as
\begin{equation}
\textbf{y} = \mathbf{W}^H\mathbf{H} \mathbf{F}_\textrm{RF} \mathbf{F}_\textrm{BB} \mathbf{s} + \mathbf{W}^H\mathbf{n},
\end{equation}
where $\mathbf{n} \in \mathbb{C}^{N_\textrm{R} \times 1} = \mathcal{C}\mathcal{N}(0,\sigma_\textrm{n}^2)$ represents the independent and identically distributed complex additive white Gaussian noise. We assume fully digital combining at the UE represented by the matrix $\mathbf{W} \in \mathbb{C}^{N_\textrm{R}\times N_\textrm{s}}$. Considering channel's singular value decomposition (SVD) as $\mathbf{H} = \mathbf{U}_\textrm{H} \mathbf{\Sigma}_\textrm{H} \mathbf{V}_\textrm{H}^H$, where $\mathbf{U}_\textrm{H} \in \mathbb{C}^{N_\textrm{R} \times N_\textrm{R}}$ and $\mathbf{V}_\textrm{H} \in \mathbb{C}^{N_\textrm{T} \times N_\textrm{T}}$ are unitary matrices, and $\mathbf{\Sigma}_\textrm{H} \in \mathbb{R}^{{N_\textrm{R} \times N_\textrm{T}}}$ is a rectangular matrix consisting of singular values as diagonal entries and rest of the entries being zero. The fully digital combiner matrix $\mathbf{W}$ is computed as the first $N_\textrm{R}$ columns of the left singular matrix $\mathbf{U}_\textrm{H}$. 

\subsection{Radar Model}
The aim of the precoding design for the radar functionality is to obtain a transmit beampattern that points to the targets of interest. For the sub-arrayed MIMO radar, the transmit beampattern can be expressed as \cite{fanTCOM2020, sellathuraiTSP2012, fanICASSP2019}
\begin{equation}\label{eq:beam_pattern}
    B(\phi) = \mathbf{a}^H_\textrm{T}(\phi^t) \mathbf{R}_\textrm{p} \mathbf{a}_\textrm{T}(\phi^t),
\end{equation}
where the covariance matrix of the precoded waveform $\mathbf{R}_\textrm{p} \in \mathbb{C}^{N_\textrm{T} \times N_\textrm{T}}$, provided $\mathbb{E}\{\mathbf{s} \mathbf{s}^H\} = \mathbf{I}_{N_\textrm{s}}$, is defined as
\begin{equation}\label{eq:covariance_matrix}
   \mathbf{R}_\textrm{p} = \mathbb{E}(\mathbf{F}_\textrm{RF} \mathbf{F}_\textrm{BB} \mathbf{s} \mathbf{s}^H \mathbf{F}_\textrm{BB}^H \mathbf{F}_\textrm{RF}^H) = \mathbf{F}_\textrm{RF} \mathbf{F}_\textrm{BB}  \mathbf{F}_\textrm{BB}^H \mathbf{F}_\textrm{RF}^H.
\end{equation}
We can observe from \eqref{eq:beam_pattern} that designing the covariance matrix in \eqref{eq:covariance_matrix} is equivalent to designing the radar beampattern. Assuming that there are $N_\textrm{p}$ targets with angular locations \{$\theta_1, \theta_2, \ldots, \theta_{N_\textrm{p}}$\}, then the diagonal elements of the optimal sub-arrayed MIMO radar-only precoder $\mathbf{F}_\textrm{RD}$ consist of $\mathbf{a}_\textrm{T}(\theta_i)\,\forall i = 1, \ldots, N_\textrm{p}$ located at the corresponding slots. The covariance matrix associated with $\mathbf{F}_\textrm{RD}$ is $\mathbf{R}_\textrm{d} = \mathbf{F}_\textrm{RD}\mathbf{F}_\textrm{RD}^H$. 

\begin{figure}[t]
\centering
    \includegraphics[width=0.45\textwidth, trim=260 120 200 150,clip]{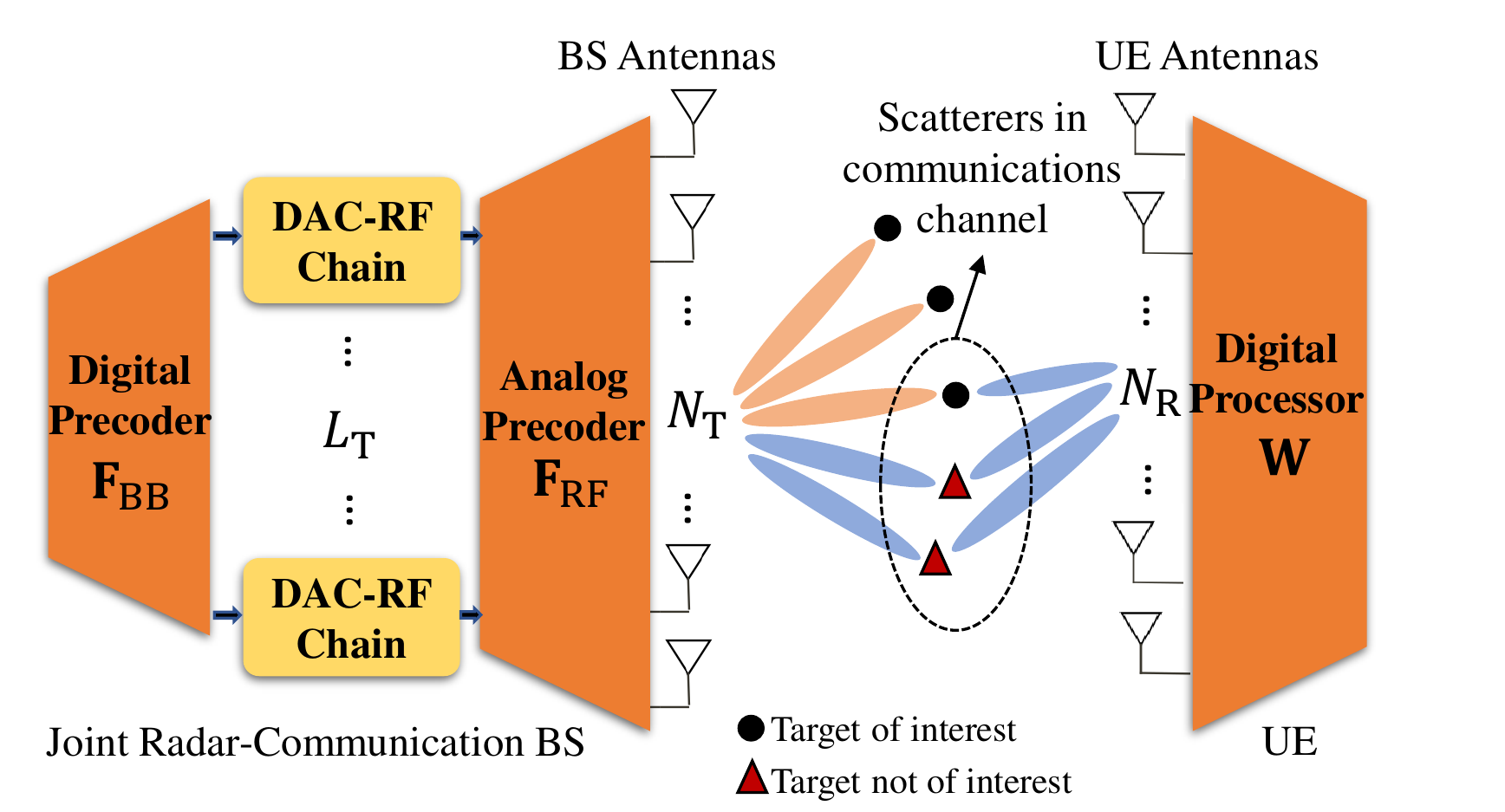}
		\caption{MIMO JRC system with hybrid precoding.}
		\vspace{-4mm}
\end{figure}

\section{EE Maximization}
We can express EE as the ratio of achievable information rate $R$ and consumed power $P$ \cite{zappone2015}:
\begin{equation}\label{eq:ee_problem}
 EE \triangleq \frac{R}{P} \,\, \textrm{(bits/Hz/J)},
\end{equation}
where the rate $R$ in bits/s/Hz can be expressed as 
\begin{align}\label{eq:rate}
R\!\triangleq &\!\log_2 \!\left \vert \mathbf{I}_{N_{\textrm{s}}}\!\!+\!\!\frac{1}{\sigma_{\textrm{n}}^2} \mathbf{W}^H\mathbf{H}\mathbf{F}_{\textrm{RF}} \mathbf{F}_{\textrm{BB}}  \mathbf{F}_{\textrm{BB}}^H  
\mathbf{F}_{\textrm{RF}}^H\mathbf{H}^H \mathbf{W} \right \vert, 
\end{align}
and the consumed power $P$ at the BS in J/s \cite{hanCM2015} is
\begin{align}\label{eq:power_bs}
\hspace{-2.5mm}P \!=\! 
a\textrm{tr}(\mathbf{F}_{\textrm{RF}}\mathbf{F}_{\textrm{BB}}\mathbf{F}_{\textrm{BB}}^H\mathbf{F}_{\textrm{RF}}^H) \!+\!L_\textrm{T}P_\textrm{RF}\!+\! N_{\textrm{T}}P_{\textrm{A}} 
\!+\!N_{\textrm{PS}}P_{\textrm{PS}} \!+\! P_{\textrm{C}}. 
\end{align}
The term $P_\textrm{RF} \!=\! 2\!\times\!(2^b P_\textrm{DAC})$ is the power per RF chain, where $b$ is the number of DAC bits associated with the RF chain unit and $P_\textrm{DAC}$ is the power per DAC bit \cite{aryanICC2019}. The term $a\textrm{tr}(\mathbf{F}_{\textrm{RF}}\mathbf{F}_{\textrm{BB}}\mathbf{F}_{\textrm{BB}}^H\mathbf{F}_{\textrm{RF}}^H)$ represents the dynamic power related to hybrid precoders where $a$ is the reciprocal of amplifier efficiency and the static power terms are as, $P_\textrm{PS}$ is the power per phase shifter, $P_\textrm{C}$ is the circuit power and $P_\textrm{A}$ is the power per antenna. The term $N_\textrm{PS}$ is the number of phase shifters, where $N_\textrm{PS}\!=\!L_\textrm{T}N_\textrm{T}$ for a fully connected structure, and $N_\textrm{PS}\!=\!N_\textrm{T}$ for a partially-connected structure \cite{vlachosVTC2018}.

We can express the EE maximization problem as follows:
\begin{align}\label{eq:ee_total}
\max_{\mathbf{F}_{\textrm{RF}}, \mathbf{F}_{\textrm{BB}}, L_{\textrm{T}}, \mathbf{U}_\textrm{T}} & \frac{R(\mathbf{F}_{\textrm{RF}}, \mathbf{F}_{\textrm{BB}}, L_{\textrm{T}})}{P(\mathbf{F}_{\textrm{RF}}, \mathbf{F}_{\textrm{BB}}, L_{\textrm{T}}) }
\nonumber \\  \textrm{subject to } &\lVert{\mathbf{F}_{\textrm{RF}}\mathbf{F}_{\textrm{BB}}\!-\!\mathbf{F}_{\textrm{RD}}\mathbf{U}_{\textrm{T}}}\rVert_F^{2} \leq \epsilon, 
\mathbf{F}_{\textrm{RF}} \!\in\! \mathcal{F}^{N_{\textrm{T}} \times L_{\textrm{T}}}, \nonumber \\
&\lVert{\mathbf{F}_{\textrm{RF}}\mathbf{F}_{\textrm{BB}}}\rVert_F^2 \!=\! P_{\textrm{max}}, 
\mathbf{U}_{\textrm{T}}\mathbf{U}_{\textrm{T}}^H \!=\! \mathbf{I}_{N_{\textrm{p}}},
\end{align}
where 
$\mathcal{F}$ is the set of possible phase shifts in $\mathbf{F}_\textrm{RF}$ and constant modulus constraints are imposed on the non-zero elements of the analog precoder, and $\epsilon$ is tolerance term. 
We multiply the desired radar precoder $\mathbf{F}_\textrm{RD}$ with unitary matrix $\mathbf{U}_\textrm{T} \in \mathbb{C}^{N_\textrm{p} \times N_\textrm{s}}$ where this auxiliary variable will not have impact on the radar beampattern \cite{fanICASSP2019}. 
The problem in \eqref{eq:ee_total} involves a fractional cost function where both the numerator and denominator parts are non-convex functions of the optimizing variables. This makes it a very difficult problem
, e.g., corresponding problem for a fully digital transceiver that admits a much simpler form is in general intractable \cite{palomarJSAC2006}. Thus for simplification, we decouple the combined problem, and in the following we discuss the RF chain optimization and hybrid precoder design.

\subsection{RF Chain Optimization}
Firstly we consider a certain available number of RF chains $L_\textrm{T}$ implemented in the hardware. 
We implement the selection mechanism between the available RF chains in the baseband domain, as part of the digital processor. In order to optimize the number of RF chains, we consider the baseband precoder matrix as $\mathbf{F}_\textrm{BB} \!=\! \mathbf{P}_\textrm{B}^{\frac{1}{2}}\hat{\mathbf{F}}_\textrm{BB}$ which inputs to the RF chain block where $\mathbf{P}_\textrm{B}$ is the diagonal matrix allocating power and $\hat{\mathbf{F}}_\textrm{BB}$ is the digital precoding matrix before the switches. 
The power allocation procedure can be described mathematically by using a diagonal sparse matrix $\mathbf{P}_\textrm{B} \in \mathcal{D}^{L_\textrm{T} \times L_\textrm{T}}$ where the set $\mathcal{D}^{L_\textrm{T} \times L_\textrm{T}} \subset \mathbb{R}^{L_\textrm{T} \times L_\textrm{T}}$ consists of ${L_\textrm{T} \times L_\textrm{T}}$ diagonal sparse matrices. To represent the baseband selection mechanism we consider that $[\mathbf{P}_\textrm{B}]_{kk} \in [0, P_\textrm{max}]$, for $k=1,\ldots, L_\textrm{T}$, where $P_\textrm{max} = \textrm{tr}(\mathbf{P}_\textrm{B}) = \sum_{k=1}^{L_\textrm{T}}[\mathbf{P}_\textrm{B}]_{kk}$. The diagonal entries of $\mathbf{P}_\textrm{B}$ with a zero value represent an open switch in the switching mechanism. Thus, the non-zero diagonal values of $\mathbf{P}_\textrm{B}$ will determine the number of the active RF chains for the BS. This switching procedure is driven by a fractional programming based approach \cite{dinkelbach1967} with the optimal power scheme to obtain the optimal number of RF chains for each channel realization. 

In order to implement the RF chain optimization procedure with power allocation based switching, we assume that $\hat{\mathbf{F}}_\textrm{BB} \hat{\mathbf{F}}_\textrm{BB}^H \approx \mathbf{I}_{L_\textrm{T}}$ \cite{ayachTWC2014}, and the matrix $\mathbf{F}_\textrm{RF}$ is defined as the $L_\textrm{T}$ columns of the $N_\textrm{T} \times N_\textrm{T}$ discrete Fourier transform (DFT) matrix.
The rate in \eqref{eq:rate} can be expressed in terms of $\mathbf{P}_\textrm{B}$ as
\begin{align}\label{eq:rate_bs}
    R(\mathbf{P}_\textrm{B}) \!=\! \log \bigg\vert \mathbf{I}_{N_\textrm{s}} \!+\! \frac{1}{\sigma_\textrm{n}^2} \mathbf{W}^H \mathbf{H} \mathbf{F}_\textrm{RF} \mathbf{P}_\textrm{B}  \mathbf{F}_\textrm{RF}^H \mathbf{H}^H \mathbf{W} \bigg\vert.
\end{align}
Similarly, we can express the power in \eqref{eq:power_bs} in terms of $\mathbf{P}_\textrm{B}$ as 
\begin{align}\label{eq:power_rf}
P(\mathbf{P}_\textrm{B}) = a\textrm{tr}(\mathbf{P}_\textrm{B}) \!+\!L_\textrm{T}^{opt}P_\textrm{RF}\!+\! N_{\textrm{T}}P_{\textrm{A}} 
\!+\!N_{\textrm{PS}}P_{\textrm{PS}} \!+\! P_{\textrm{C}},
\end{align}
where $L_\textrm{T}^{opt}$ is the optimal number of RF chains that needs to be computed using the proposed method.
Note that in order to obtain optimal number of RF chains, the EE maximization in \eqref{eq:ee_total} is considered only for the communication scenario while it is expressed into a weighted summation for both radar and communication operations when computing hybrid precoders. The EE maximization following rate and power in \eqref{eq:rate_bs}-\eqref{eq:power_rf} is 
\begin{equation}\label{eq:ee_constraint}
\max_{\mathbf{P}_\textrm{B} \in \mathcal{D}^{L_\textrm{T} \times L_\textrm{T}}}\! \frac{R(\mathbf{P}_\textrm{B})}{P(\mathbf{P}_\textrm{B})} \textrm{ s. t. }  P(\mathbf{P}_\textrm{B})\!\leq\! \hat{P}_{\textrm{max}}, R(\mathbf{P}_\textrm{B})\!\geq\!R_{\textrm{min}},
\end{equation}
where the first constraint term sets the upper bound for the total power budget, i.e.,
$\hat{P}_\textrm{max} \!=\! P_\textrm{max} \!+\! L_\textrm{T}P_\textrm{RF}\!+\! N_\textrm{T} P_\textrm{A} \!+\! N_\textrm{PS}P_\textrm{PS} + P_\textrm{C}$, and $R_\textrm{min}$ represents the minimum achievable rate.

\begin{algorithm}[t]
\begin{algorithmic}[1]
\STATE \textbf{Given} $\mathbf{P}_\textrm{B}^{(0)}, \nu^{(0)}$, $\mathcal{G}(\mathbf{P}_\textrm{B}^{(0)},\nu^{(0)}) \geq 0$, $m = 0$, $\beta$, $L_\textrm{T}$  \\
\STATE \textbf{while} $| \mathcal{G}(\mathbf{P}_\textrm{B}^{(m)}, \nu^{(m)})| > \beta$ \textbf{do}\\
\STATE \hspace{2mm} Solve \eqref{eq:db_step} by alleviating constraint on $\mathbf{P}_\textrm{B}^{(m)}$ \\
\STATE \hspace{2mm} Threshold $\mathbf{P}_\textrm{B}^{(m)}$ for non-zero values and update $L_\textrm{T}^{opt}$ \\
\STATE \hspace{2mm} Compute \eqref{eq:rate_approximation}, \eqref{eq:p_tx_over_p} and $\mathcal{G}(\mathbf{P}^{(m)}, \nu^{(m)})$  
\STATE \hspace{2mm} Update $\nu^{(m)}$ value as $R(\mathbf{P}_\textrm{B}^{(m)})/P(\mathbf{P}_\textrm{B}^{(m)})$\\
\STATE \hspace{2mm} Update $m$-th to next $(m+1)$-th iteration\\
\STATE \textbf{end while}\\
\STATE Obtain $L_\textrm{T}^{opt}$ as L$0$-norm of thresholded $\mathbf{P}_\textrm{B}^{(m)}$
\end{algorithmic}
\caption{Obtaining Optimal Number of RF Chains}
\end{algorithm}

Using fractional programming approach \cite{dinkelbach1967}, we can express \eqref{eq:ee_constraint} with iterative difference-based optimizations as
\begin{align}\label{eq:ee_constraint2}
&\max_{\mathbf{P}_\textrm{B}^{(m)} \in \mathcal{D}^{L_\textrm{T} \times L_\textrm{T}}} \left\{ R(\mathbf{P}_\textrm{B}^{(m)}) - \nu^{(m)} P(\mathbf{P}_\textrm{B}^{(m)}) \right\} 
\nonumber \\
&\textrm{ subject to } P(\mathbf{P}_\textrm{B}) \leq \hat{P}_{\textrm{max}}, R(\mathbf{P}_\textrm{B}) \geq R_{\textrm{min}}.
\end{align}
This approach involves a sequence of iterations where the constant $\nu^{(m)}$ is updated at each iteration based on the rate and power values estimated during the previous iteration, i.e., $R(\mathbf{P}_\textrm{B}^{(m\!-\!1)})/P(\mathbf{P}_\textrm{B}^{(m\!-\!1)}) \!\in\! \mathbb{R}^+\,\forall\, m\!=\!\!1,2,..,M_{\textrm{max}}$, where $M_{\textrm{max}}$ denotes the maximum number of iterations.  We can solve \eqref{eq:ee_constraint2} with respect to (w.r.t.) $\mathbf{P}_\textrm{B}$ using Dinkelbach algorithm \cite{dinkelbach1967}.

Given that $\mathbf{W}\mathbf{W}^H = \mathbf{I}_{N_\textrm{s}}$, and for large number of antenna arrays, it can be observed that $\mathbf{\Psi} \triangleq \mathbf{W}^H \mathbf{H} \mathbf{F}_\textrm{RF} \mathbf{F}_\textrm{RF}^H \mathbf{H}^H \mathbf{W}$ is a strongly-diagonal matrix. Then, we approximate \eqref{eq:rate_bs} as
\begin{equation}\label{eq:rate_tx_svd}
R(\mathbf{P}_\textrm{B}) = \log \bigg\vert \mathbf{I}_{N_\textrm{s}} + \frac{1}{\sigma_\textrm{n}^2} \mathbf{\hat{\Sigma}}^2 \mathbf{P}_\textrm{B} \bigg\vert,
\end{equation}
where $\mathbf{\hat{\Sigma}}$ is the diagonal matrix with the singular values of $\mathbf{\Psi}$. As \eqref{eq:rate_tx_svd} involves only diagonal matrices, we can decompose the rate into $L_\textrm{T}$ parallel and orthogonal channels as
\begin{align}\label{eq:rate_approximation}
    R(\mathbf{P}_\textrm{B}) \approx  \sum_{k=1}^{L_\textrm{T}} \log \left(1 +  \frac{1}{\sigma_\textrm{n}^2} [\mathbf{\hat{\Sigma}}^2]_{kk} [\mathbf{P}_\textrm{B}]_{kk}\right).
\end{align}
Also, the power in terms of $\mathbf{P}_\textrm{B}$ can be expressed as
\begin{align}
    P(\mathbf{P}_\textrm{B}) &= \sum_{k=1}^{L_\textrm{T}} \alpha [\mathbf{P}_\textrm{B}]_{kk} + \mathbf{P}_\textrm{S}, \label{eq:p_tx_over_p}
\end{align}
where $\mathbf{P}_\textrm{S}$ denotes the static power term which is independent of $\mathbf{P}_\textrm{B}$. Considering $\sum_{k=1}^{L_{\textrm{T}}} [\mathbf{P}_\textrm{B}]_{kk} = \textrm{tr}(\mathbf{P}_\textrm{B}) = P_\textrm{max}$, for a partially-connected structure we have $\mathbf{P}_\textrm{S} = N_{\textrm{T}}P_{\textrm{A}}+N_\textrm{T}P_\textrm{PS} + P_{\textrm{C}}$ and $\alpha \triangleq a + \frac{P_{\textrm{RF}}}{P_\textrm{max}}$, whereas for a fully-connected structure $\mathbf{P}_\textrm{S} = N_{\textrm{T}}P_{\textrm{A}}+ P_{\textrm{C}}$ and $\alpha \triangleq a + \frac{P_{\textrm{RF}}+N_\textrm{T}P_\textrm{PS}}{P_\textrm{max}}$. For \eqref{eq:rate_approximation} and \eqref{eq:p_tx_over_p}, $m$-th step of the proposed method can be computed as
\begin{align}
&\{ \mathbf{P}_\textrm{B}^{(m)}, \nu^{(m)} \} = \textrm{arg} \max_{\mathbf{P}_\textrm{B}^{(m)} \in \mathcal{D}^{L_\textrm{T} \times L_\textrm{T}}} \mathcal{G}(\mathbf{P}_\textrm{B}^{(m)} \nu^{(m)}), \nonumber \\ &\textrm{ subject to } \, P(\mathbf{P}_\textrm{B}) \leq \hat{P}_{\textrm{max}}, R(\mathbf{P}_\textrm{B}) \geq R_{\textrm{min}}, \label{eq:db_step}
\end{align}
\text{where} $\mathcal{G}(\mathbf{P}_\textrm{B}^{(m)}, \nu^{(m)}) \triangleq \sum_{k=1}^{L_\textrm{T}}  \log \left(1 +  \frac{1}{\sigma_\textrm{n}^2} [\boldsymbol{\hat{\Sigma}}^2]_{kk} [\mathbf{P}_\textrm{B}^{(m)}]_{kk}\right) 
- \nu^{(m)}  \sum_{k=1}^{L_\textrm{T}} \alpha [\mathbf{P}_\textrm{B}^{(m)}]_{kk}$. This problem in \eqref{eq:db_step} is non convex due to constraint on $\mathbf{P}_\textrm{B}^{(m)}$. Algorithm 1 states the procedure to obtain the optimal number of RF chains using Dinkelbach method. Note that in Step 3, first we alleviate the constraint on $\mathbf{P}_\textrm{B}^{(m)}$ in order to solve \eqref{eq:db_step} and then in Step 4, we impose thresholding where all the values below a tolerance value $\beta_1$ become zero and the non-zero entries determine the number of RF chains to be activated, i.e., $L_\textrm{T}^{opt}$. 
In this method, we use rate and power that do not depend explicitly on $\mathbf{F}_\textrm{RF}$ and $\mathbf{F}_\textrm{BB}$, which avoids the need to compute these matrices each time the number of active RF chains to be selected, thus this method provides a low complexity solution. Reference \cite{aryanTGCN2019} can be referred for further insights on the computational complexity of the iterative procedure. Once we obtain the optimal number of RF chains, hybrid precoder matrices can be designed based on a weighted minimization problem. 



\subsection{Hybrid Precoder Design}
In order to design the hybrid precoders at the BS, the fractional problem of EE maximization in \eqref{eq:ee_total} can be expressed into an equivalent non-fractional minimization problem using fractional programming approach in \cite{dinkelbach1967}. Following the simplified problem, the optimal precoding matrices can be designed such that the mutual information achieved by Gaussian signaling over the wireless channel is maximized \cite{ayachTWC2014}. We adopt the approach in \cite{ayachTWC2014} where the maximization of the mutual information can be approximated by finding the minimum Euclidean distance of the hybrid precoder to the fully digital transceiver. For example, \cite{aryanTGCN2020} discusses steps to transform a difficult fractional EE maximization problem into Euclidean distance minimization problem. 

Furthermore, considering weighted summation for radar and communication operations, it can be assumed that the decomposition $\mathbf{F}_\textrm{RF}\mathbf{F}_\textrm{BB}$ can be made sufficiently close to the optimal fully digital precoding matrix $\textbf{F}_\textrm{DF}$ during the communication operation and simultaneously close to the optimal radar precoder matrix $\mathbf{F}_\textrm{RD}\mathbf{U}_\textrm{T}$ for the radar operation \cite{fanICASSP2019}. 
The joint radar-communications based weighted summation problem for the hybrid precoder design can be expressed as
\begin{align}\label{eq:precoders}
&\underset{\mathbf{F}_\textrm{RF}, \mathbf{F}_\textrm{BB},\mathbf{U}_\textrm{T}}{\textrm{min}}
\rho \!\norm{\mathbf{F}_\textrm{RF}\mathbf{F}_\textrm{BB}\!-\!\!\mathbf{F}_\textrm{DF}}_F^{2} \!+\!\! (1\!\!-\!\rho) \norm{\mathbf{F}_\textrm{RF}\mathbf{F}_\textrm{BB}\!-\!\!\mathbf{F}_\textrm{RD}\mathbf{U}_\textrm{T} }_F^{2}
\nonumber \\
&\textrm{s. t. }
\mathbf{F}_\textrm{RF} \!\in\! \mathcal{F}^{N_{\textrm{T}} \times L_{\textrm{T}}}, 
\norm{\mathbf{F}_\textrm{RF}\mathbf{F}_\textrm{BB}}_F^2\!=\!P_{\textrm{max}}, \mathbf{U}_\textrm{T}\mathbf{U}_\textrm{T}^H\!=\!\mathbf{I}_{N_\textrm{p}},
\end{align}
where $\rho \in [0,1]$ is the weighting factor used in determining the weights for radar and communication operations. The term $(1-\rho)$ acts as the Lagrangian multiplier for the first constraint of \eqref{eq:ee_total} which sets equivalence between \eqref{eq:ee_total} and \eqref{eq:precoders}. Note that, unlike the RF chain optimization procedure in Subsection III.A, the matrices $\mathbf{F}_\textrm{RF}$ and $\mathbf{F}_\textrm{BB}$ are not predefined but obtained via the solution of \eqref{eq:precoders}. Also the assumption of orthonomality on the digital precoder matrix may not be followed. A high value of $\rho$ places priority on the communication operation and a low $\rho$-value prioritises radar functionality. Regarding the transmit power constraint $P_\textrm{max}$ for the hybrid precoder decomposition in \eqref{eq:precoders}, equality is taken into consideration for practical scenario as radar operation may require to transmit signal at maximum available power \cite{stoicaTSP2007}. Equation \eqref{eq:precoders} involves non-convexity in the objective function and constraints, which makes the problem difficult. 

To solve \eqref{eq:precoders}, we use an efficient and iterative alternating minimization based approach to achieve near optimal solution. 
We decompose \eqref{eq:precoders} into three separate sub-problems where each one computes these matrices individually. When $\mathbf{F}_\textrm{RF}$ and $\mathbf{F}_\textrm{BB}$ are fixed, \eqref{eq:precoders} can be formulated as \cite{fanICASSP2019}
\begin{align}\label{eq:UT_matrix}
\underset{\mathbf{U}_\textrm{T}}{\textrm{min}}
\norm{\mathbf{F}_\textrm{RD}\mathbf{U}_\textrm{T}\!-\!\! \mathbf{F}_\textrm{RF}\mathbf{F}_\textrm{BB}}_F^{2}
\textrm{ subject to } \mathbf{U}_\textrm{T}\mathbf{U}_\textrm{T}^H = \mathbf{I}_{N_\textrm{p}},
\end{align}
which can be solved via SVD as $\mathbf{U}_\textrm{T} = \hat{\mathbf{U}}_\textrm{T}\mathbf{I}_{N_\textrm{p}\times N_\textrm{s}}\hat{\mathbf{V}}_\textrm{T}$, where $\hat{\mathbf{U}}_\textrm{T}\boldsymbol{\Sigma}_\textrm{T}\hat{\mathbf{V}}_\textrm{T}=\mathbf{F}_\textrm{RD}^H\mathbf{F}_\textrm{RF}\mathbf{F}_\textrm{BB}$ is the SVD of $\mathbf{F}_\textrm{RD}^H\mathbf{F}_\textrm{RF}\mathbf{F}_\textrm{BB}$. When $\mathbf{U}_\textrm{T}$ and $\mathbf{F}_\textrm{BB}$ are fixed, \eqref{eq:precoders} can be formulated as
\begin{align}\label{eq:FRF_matrix}
\underset{\mathbf{F}_\textrm{RF}}{\textrm{min}}\,&
\rho\norm{\mathbf{F}_\textrm{RF}\mathbf{F}_\textrm{BB}\!-\!\!\mathbf{F}_\textrm{DF}}_F^2+(1-\rho)\norm{\mathbf{F}_\textrm{RF}\mathbf{F}_\textrm{BB}\!-\!\!\mathbf{F}_\textrm{RD}\mathbf{U}_\textrm{T}}_F^2
\nonumber \\ 
&\textrm{ subject to } \mathbf{F}_\textrm{RF} \in \mathcal{F}^{N_{\textrm{T}} \times L_{\textrm{T}}},
\end{align}
which can be solved for each non-zero element of $\mathbf{F}_\textrm{RF}$ and an optimal solution to equivalent phase rotation problem can be obtained, such as $[\mathbf{F}_\textrm{RF}]_{kl} = \textrm{exp}(j\,\textrm{arg}\,\{\mathbf{p}^H\mathbf{q}\})$, where $\mathbf{p}=[\sqrt{\rho}[\mathbf{F}_\textrm{DF}]_{k,:},\sqrt{1-\rho}[\mathbf{F}_\textrm{RD}\mathbf{U}_\textrm{T}]_{k,;}]^T$ and $\mathbf{q}=[\sqrt{\rho}[\mathbf{F}_\textrm{BB}]_{l,:},\sqrt{1-\rho}[\mathbf{F}_\textrm{BB}]_{k,;}]^T$ \cite{fanICASSP2019}. Lastly, when $\mathbf{U}_\textrm{T}$ and $\mathbf{F}_\textrm{RF}$ are fixed, \eqref{eq:precoders} can be formulated as
\begin{align}\label{eq:FBB_matrix}
\underset{\mathbf{F}_\textrm{BB}}{\textrm{min}}\,&
\rho\norm{\mathbf{F}_\textrm{RF}\mathbf{F}_\textrm{BB}\!-\!\!\mathbf{F}_\textrm{DF}}_F^2+(1-\rho)\norm{\mathbf{F}_\textrm{RF}\mathbf{F}_\textrm{BB}\!-\!\!\mathbf{F}_\textrm{RD}\mathbf{U}_\textrm{T}}_F^2
\nonumber \\ 
&\textrm{ subject to } \norm{\mathbf{F}_\textrm{BB}}_F^2=L_\textrm{T}P_\textrm{max}/N_\textrm{T},
\end{align}
where as per the special structure of $\mathbf{F}_\textrm{RF}$, the power constraint is recast as $\norm{\mathbf{F}_\textrm{RF}\mathbf{F}_\textrm{BB}}_F^2=N_\textrm{T}/L_\textrm{T}\norm{\mathbf{F}_\textrm{BB}}_F^2=P_\textrm{max}$. Following formulation of \eqref{eq:FBB_matrix} shown in \cite{fanICASSP2019} and involving eigen-value decomposition and golden-section search steps shown in \cite{fanTSP2018_2}, the solution to $\mathbf{F}_\textrm{BB}$ can be obtained. Algorithm 2 summarizes the steps to obtain optimal $\mathbf{U}_\textrm{T}$, $\mathbf{F}_\textrm{RF}$ and $\mathbf{F}_\textrm{BB}$ matrices.
Once the optimal number of RF chains and hybrid precoder matrices are obtained, rate in \eqref{eq:rate} and power in \eqref{eq:power_bs} are computed resulting into achieving maximum EE in \eqref{eq:ee_problem}. 

\begin{algorithm}[t]
\begin{algorithmic}[1]
\STATE \textbf{Given} $\mathbf{F}_\textrm{DF}$, $\mathbf{F}_\textrm{RD}$, $\mathbf{H}$, $\rho$, $\beta_2\!>\!0$, $n\!=\!\!1$, total iterations $N_\textrm{max}$  \\
\STATE Initializing $\mathbf{U}_\textrm{T}^{(0)}$, $\mathbf{F}_\textrm{RF}^{(0)}$, $\mathbf{F}_\textrm{BB}^{(0)}$ randomly and denoting the optimization in $\eqref{eq:precoders}$ as function $g^{(0)}$ 
\STATE \textbf{while} $n\leq N_\textrm{max}$ and $|g^{(n)}-g^{(n-1)}|\geq \beta_2$ \textbf{do}\\
\STATE \hspace{2mm} Solve sub-problem \eqref{eq:UT_matrix} to compute $\mathbf{U}_\textrm{T}^{(n)}$ \\
\STATE \hspace{2mm} Solve sub-problem \eqref{eq:FRF_matrix} to compute $\mathbf{F}_\textrm{RF}^{(n)}$ \\
\STATE \hspace{2mm} Solve sub-problem \eqref{eq:FBB_matrix} to compute $\mathbf{F}_\textrm{BB}^{(n)}$ \\ 
\STATE \hspace{2mm} Compute objective function $g^{(k)}$ using $\mathbf{U}_\textrm{T}^{(n)}$, $\mathbf{F}_\textrm{RF}^{(n)}$, $\mathbf{F}_\textrm{BB}^{(n)}$ \\
\STATE \hspace{2mm} Update $n$-th to next $(n+1)$-th iteration\\
\STATE \textbf{end while}\\
\STATE Obtain optimal $\mathbf{U}_\textrm{T}$, $\mathbf{F}_\textrm{RF}$, $\mathbf{F}_\textrm{BB}$ matrices
\end{algorithmic}
\caption{Hybrid Precoder Computation}
\end{algorithm}

\vspace{-0.7mm}
\section{Simulation Results}
We evaluate the performance of the proposed method where results are averaged over 1,000 Monte-Carlo realizations.
\subsubsection*{System setup}
We set system parameter values as $N_\textrm{T} = 120$, $N_\textrm{R} = 6$, $N_\textrm{s} = 6$, $N_\textrm{p} = 3$ and $N_\textrm{m} = 10$. For the power values, we follow \cite{aryanTGCN2020} and references therein, and set $P_\textrm{DAC} = 1$ mW, $P_\textrm{PS} = 10$ mW, $P_\textrm{A} = 100$ mW and $P_\textrm{C} = 10$ W. We set the three targets to be located at $[-30^\circ, 0^\circ, 30^\circ]$. We follow standard assumptions for the channel gain $\alpha_l$ being subjected to standard complex Gaussian distribution, and $\phi_l^t$ and $\phi_l^r$ follow the uniform distribution in $[-180^\circ, 180^\circ]$. The available number of RF chains $L_\textrm{T}$ is the length of the eigenvalues obtained from $\boldsymbol{\Psi}$. The fully digital precoder $\mathbf{F}_\textrm{DF}$ is
computed as the first $N_\textrm{R}$ columns of the matrix $\mathbf{V}_\textrm{H}$.
For $P_\textrm{RF}$ in \eqref{eq:power_bs}, we consider full-bit DAC resolution, thus $b=8$ \cite{aryanTGCN2020}.  The signal to noise ratio (SNR) is $1/\sigma_\textrm{n}^2$, $R_\textrm{min}\!=\!\!1$ bits/s/Hz, $P_\textrm{max}\!=\!1$ W, and tolerance values $\beta\!=\!\!10^{-4}$ and $\beta_1\!=\!\!10^{-6}$.


\subsubsection*{Performance evaluation} Fig. 2 shows the EE and rate performance w.r.t. SNR for fixed $N_\textrm{R} = 6$. The weighting factor $\rho$ is varied from 0.4 to 1 and for $\rho= 1$, communication only operation takes place. In these plots, the performance of the proposed method is compared with the optimal fully digital precoder also. It can be observed that with increase in $\rho$, the EE and rate performance increases as higher weight is allocated to obtain communication hybrid precoder close to the optimal fully digital precoder, e.g., at 5 dB SNR, EE for $\rho=1$ is $\approx$ 0.15 times better than the EE for $\rho=0.4$. The proposed method achieves better EE than the optimal fully digital precoder for different $\rho$ values where best performance is achieved for $\rho = 1$, e.g., at 5 dB SNR, EE for $\rho = 1$ is $\approx$ 0.25 times better than the optimal fully digital precoder, and $\rho=0.4$ outperforms fully digital precoder by $\approx$ 0.1 times. 
Furthermore, the proposed method achieves good rate performance close to the optimal fully digital precoder, specially for low SNR region, e.g., at -25 dB SNR, rate performance of the hybrid precoder design for different $\rho$ values is similar to the optimal fully digital precoder. 

\begin{figure}[t]
 \centering 
 \includegraphics[width=0.51\textwidth, trim=65 0 0 0,clip]{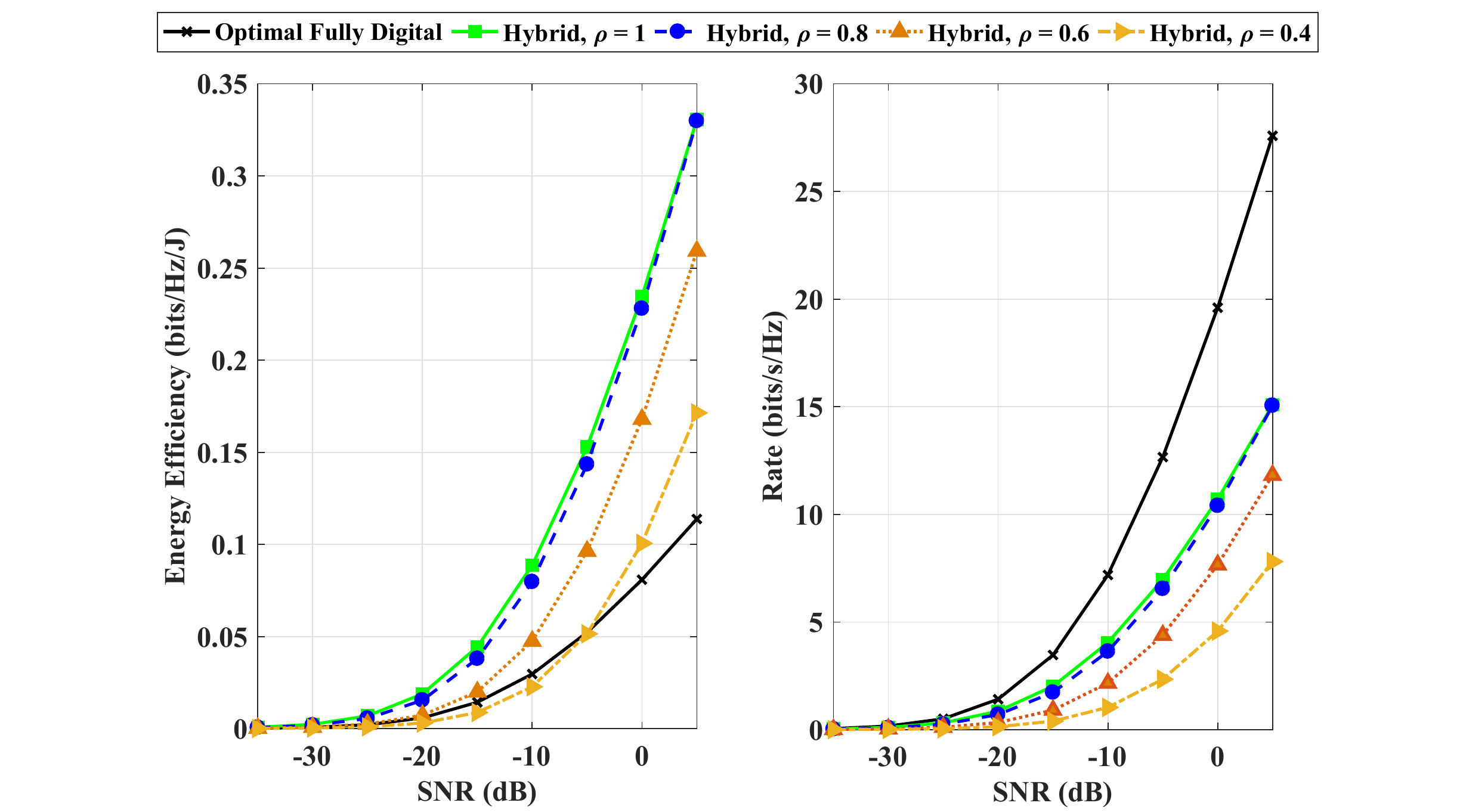}
 		\caption{EE and rate communication performance w.r.t. SNR for $N_{\textrm{T}}\!=\!120$, $N_{\textrm{R}}\!=6, N_{\textrm{s}} \!=\!6$ and $N_{\textrm{p}}\!=\!3$.}
 \end{figure}
 
 \begin{figure}[t]
 \centering 
 \includegraphics[width=0.465\textwidth, trim=50 0 10 0,clip]{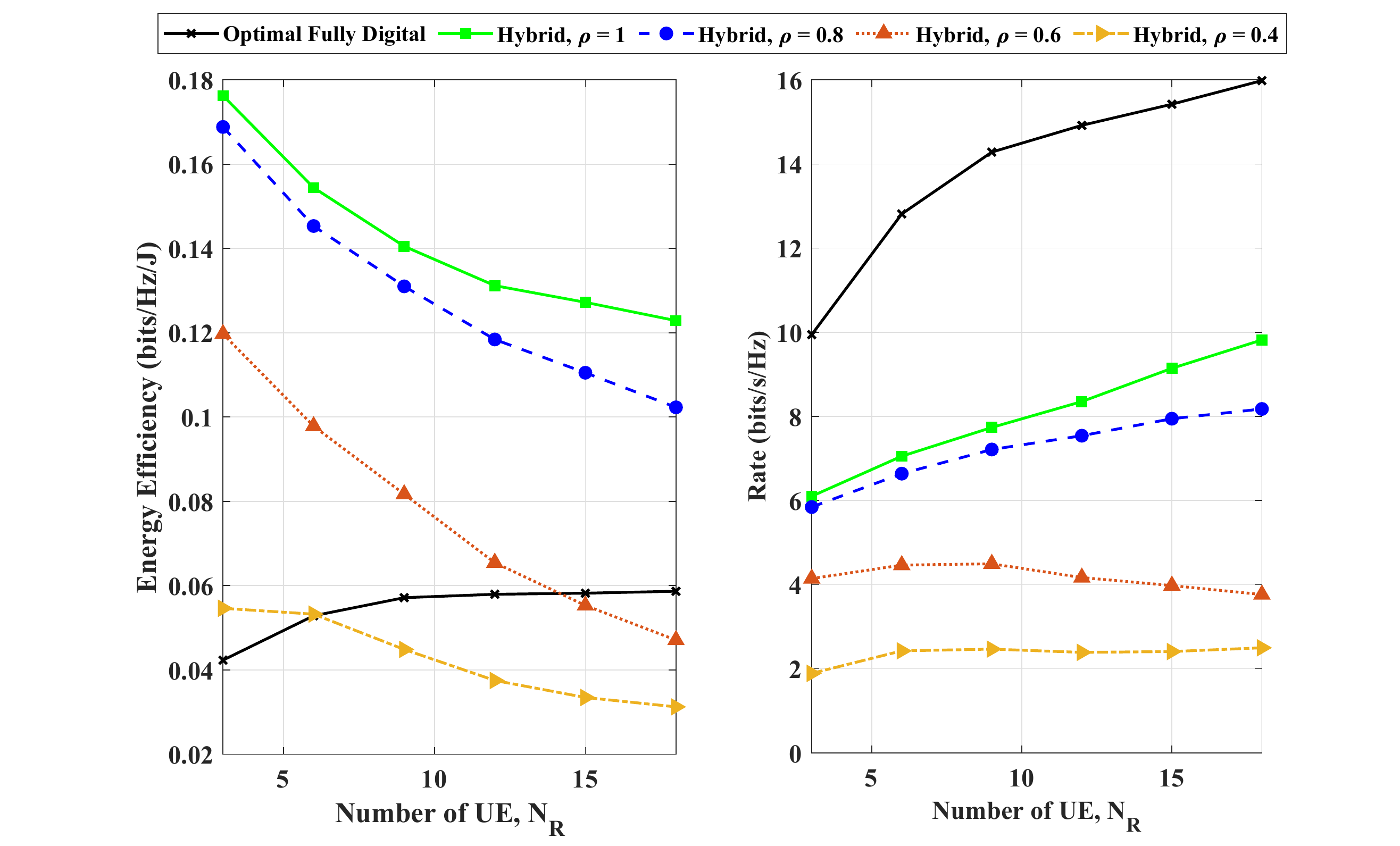}
 		\caption{EE and rate communication performance w.r.t. $N_\textrm{R}$ for SNR $\!=-5$ dB, $N_{\textrm{T}}\!=\!120$, $N_{\textrm{s}} \!=\!6$ and $N_{\textrm{p}}\!=\!3$.}
 \end{figure}

\begin{figure}[t]
 \centering 
 \includegraphics[width=0.5\textwidth, trim=10 5 0 0,clip]{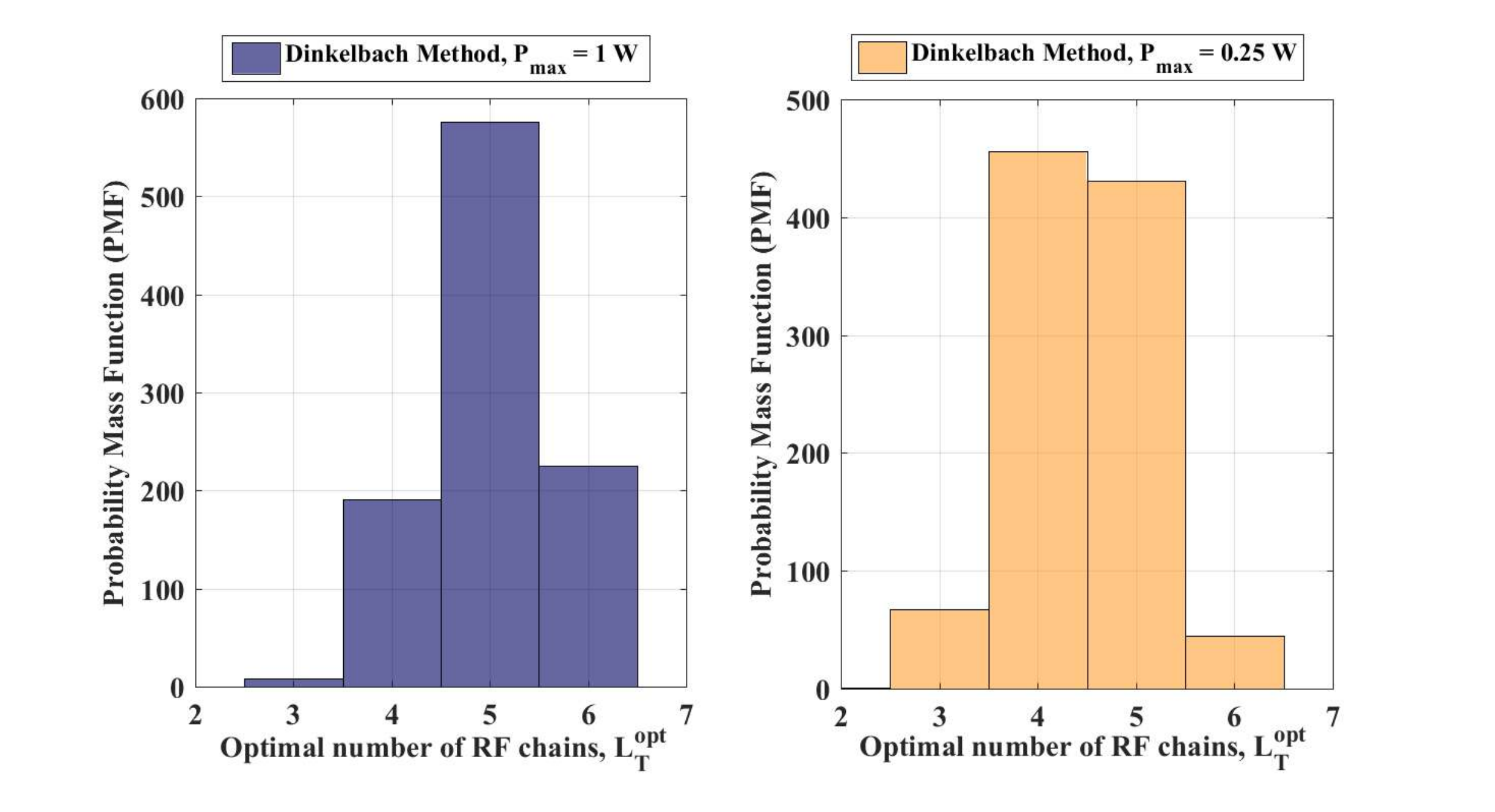}
 		\caption{PMF plots for the proposed RF chain selection method for $N_\textrm{T} = 120$ at $P_\textrm{max} = 1$ W and $P_\textrm{max} = 0.25$ W.}
 \end{figure}

Fig. 3 shows the EE and rate performance w.r.t. $N_\textrm{R}$ for fixed SNR=-5 dB. Similarly to Fig. 2, the EE and rate performance for the hybrid precoder design with optimal RF chain selection is higher with high value of weighting factor $\alpha$, e.g., at $N_\textrm{R}=3$, EE for $\rho=1$ is $\approx$ 0.12 times better than EE for $\rho = 0.4$ and rate for $\rho=1$ is $\approx$ 4 bits/s/Hz better than rate for $\rho=0.4$. It can be observed that the proposed method achieves best EE performance for $\rho=1$ and better than the fully digital precoder, e.g., at $N_\textrm{R}=3$, EE for $\rho=1$ outperforms EE for fully digital precoder by $\approx$ 0.14 times.

Fig. 4 shows the significance of the proposed RF chain selection method in terms of the probability mass function (PMF). 
We consider the PMF distribution of the proposed Dinkelbach method over the selected optimal number of RF chains, i.e., $L_\textrm{T}^{opt}$, for $P_\textrm{max} = 1$ W and $P_\textrm{max} = 0.25$ W. PMF means that for how many channel realizations, the proposed Dinkelbach method finds a particular optimal number of RF chains out of $L_\textrm{T}$ available RF chains at different values of $P_\textrm{max}$. We can observe that the proposed method is highly hardware efficient as, at $P_\textrm{max} = 1$ W, for most number of channel realizations, i.e., $\approx{580}$ realizations, it activates only $5$ RF chains for $120$ transmit antennas, and similarly at $P_\textrm{max} = 0.25$ W, it activates only $4$ RF chains for $\approx{450}$ realizations and $5$ RF chains for $\approx{420}$ realizations. 


\begin{figure}[t]
 \centering 
 \includegraphics[width=0.52\textwidth, trim=20 0 0 0,clip]{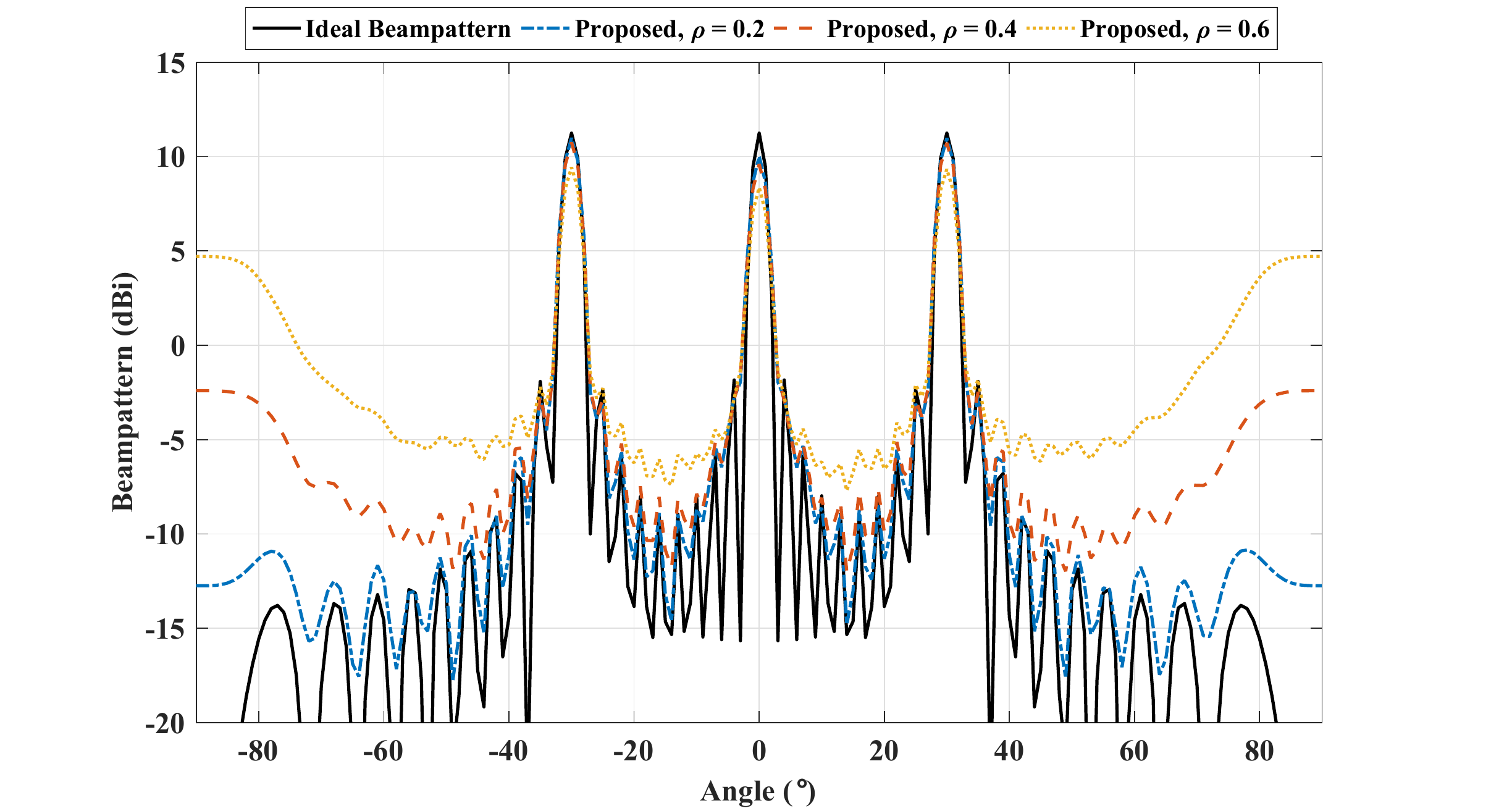}
 		\caption{Radar beampattern performance at different $\rho$ values for $N_{\textrm{T}}\!=\!120$, $N_{\textrm{R}}\!=6, N_{\textrm{s}} \!=\!6$ \& $N_{\textrm{p}}\!=\!3$.}
 		\vspace{-2mm}
 \end{figure}

Fig. 5 shows the radar beampattern performance for the proposed approach with different $\rho$ values, which is compared with the ideal radar beampattern. We can observe that the BS can effectively steer beams in the direction of targets while maintaining reasonable communication performance. For instance, at $\rho = 0.2$ and $\rho = 0.4$ values when radar operation is dominating, radar beampattern associated with the proposed approach is similar to the range mainlobes of ideal beampattern and approaches close to the range sidelobes. A trade-off between sensing and communication can be observed for higher $\rho$ value when communication operation dominates, e.g., at $\rho = 0.6$, where the communication transmission gives rise to the sidelobes observed at $+-90^{\circ}$.

\section{Conclusion}
This paper proposes a joint sensing and communication by designing energy efficient JRC system with minimum hardware requirement. A dynamic selection mechanism selects the optimal number of RF chains at current channel state and near optimal hybrid precoders are designed with weighted formulation of the communication and radar operations. The proposed method saves hardware complexity by only activating the required number of RF chains and achieves a scalable trade-off between communications and radar performance. 
The proposed method is highly hardware efficient, e.g., it activates only $5$ RF chains for massive $120$ transmit antennas. An efficient trade-off between sensing and downlink communication is achieved, and we obtain favourable radar beampattern performance. 

\section*{Acknowledgment}
This work was supported by the Engineering and Physical Sciences Research Council of the UK (EPSRC) Grant number EP/S026622/1 and the UK MOD University Defence Research Collaboration (UDRC) in Signal Processing.

\bibliographystyle{IEEEtran}

\end{document}